\documentclass[conference]{IEEEtran}

\usepackage{amsmath,amssymb}
\usepackage{booktabs}
\usepackage{cite}
\usepackage{graphicx}
\usepackage{multirow}
\usepackage{xcolor}
\usepackage{tikz}
\usepackage{url}
\usetikzlibrary{arrows.meta,fit,positioning}
\usepackage{soul}
\usepackage{subcaption}
\usepackage{microtype}
\newif\ifshowcomments
 \showcommentsfalse

\ifshowcomments
  \newcommand{\td}[1]{\sethlcolor{pink}\hl{[Ting: #1]}}
  
  \definecolor{myblue}{RGB}{200,220,235}
  \newcommand{\as}[1]{\sethlcolor{myblue}\hl{[Austin: #1]}}
  \definecolor{mygreen}{RGB}{210,235,210}
  \newcommand{\ys}[1]{\sethlcolor{mygreen}\hl{[Yishun: #1]}}
  \newcommand{\ysrev}[1]{\textcolor{blue}{#1}}
\else
  \newcommand{\td}[1]{}
  \newcommand{\as}[1]{}
  \newcommand{\ys}[1]{}
  
  \newcommand{\ysrev}[1]{#1}
\fi
\newcommand{\gain}[1]{\textbf{#1}}

\title{QuaSR: Quality-Aware Sample Reweighting for Pacific Indigenous Speech Recognition }

\author{
    \IEEEauthorblockN{
        Yishun Li\IEEEauthorrefmark{1}, 
        Yang Xiao\IEEEauthorrefmark{1}, 
        Gongping Huang\IEEEauthorrefmark{2}, 
        Eun-Jung Holden\IEEEauthorrefmark{1}, 
        Nick Thieberger\IEEEauthorrefmark{1}, 
        Ting Dang\IEEEauthorrefmark{1}
    }
    
    \IEEEauthorblockA{
        \IEEEauthorrefmark{1}The University of Melbourne, Australia
    }
    \IEEEauthorblockA{
        \IEEEauthorrefmark{2}Wuhan University, China
    }

}

\begin{document}
\maketitle

\begin{abstract}

Training automatic speech recognition (ASR) models for low-resource languages is challenging due to limited data and highly variable supervision quality. In particular, Pacific Indigenous speech corpora often exhibit heterogeneous acoustic conditions, transcript inconsistencies, and varying degrees of acoustic–text alignment reliability, making standard fine-tuning approaches sensitive to noisy or misleading supervision signals. In this work, we propose QuaSR, a simple yet effective weighting framework that combines data-side reliability with model-side learnability to improve ASR adaptation. Specifically, we estimate data reliability from acoustic, transcription, and alignment, while measuring learnability using training loss from the model. These two complementary signals are integrated into a unified sample utility score to produce training weights for the samples. We also evaluated across four Pacific Indigenous languages, which shows that the proposed utility scores reliably correlate with adaptation performance. Furthermore, QuaSR consistently improves ASR adaptation over standard fine-tuning and alternative data selection strategies, highlighting a new way to leverage difficulty scores for low-resource speech learning.
\end{abstract}


\section{Introduction}

Automatic speech recognition (ASR) has achieved substantial improvements on high-resource language benchmarks in recent years \cite{radford2022whisper,pratap2023mms,yadav2022survey}. However, these advances remain unevenly distributed across the world's languages. For many low-resource languages, the primary bottleneck is not model capacity but the limited availability of suitable speech data for adaptation and evaluation. Collecting and annotating speech corpora is costly and time-consuming, and available recordings are often sourced from linguistic documentation projects, field recordings, or archival collections rather than speech datasets specifically designed for ASR \cite{leferrand2025field}. Pacific Indigenous languages exemplify this challenge: available data are often limited in scale, heterogeneous in recording conditions, and inconsistent in transcription quality and segmentation. Consequently, these resources are often poorly matched to the assumptions of modern end-to-end ASR systems, which generally perform best with large quantities of consistently segmented and transcribed speech \cite{xiao2026continual}. 
Due to the resource challenge, it is difficult to train an ASR model from scratch for Pacific Indigenous languages.

With the notable zero-shot performance with speech foundation models (SFM), recent work has begun exploring parameter-efficient fine-tuning (PEFT) methods to adapt SFM to low-resource languages. Whilst these approaches have demonstrated a degree of effectiveness, performance on Pacific Indigenous languages remains consistently below that of better-resourced languages, and varies considerably even across Pacific languages themselves \cite{xiao2026continual,xiao2026dama}. This disparity is most likely attributable to the extremely limited data volume available for these languages and, more critically, to fundamental issues of data quality 
when collecting the Pacific Indigenous languages data.

Collecting speech data from Pacific Indigenous communities presents significant practical challenges. Recording conditions are rarely controlled, resulting in substantial acoustic variability and background noise. Transcription poses an equally serious obstacle: accurate annotation requires access to community language experts, who are few in number and whose availability is often constrained \cite{leferrand2025field,shi2021leveraging,xiao2022continual,xiao2022rainbow}. Consequently, transcriptions may be incomplete, inconsistent, or of uncertain reliability, a problem that is compounded when the underlying recordings are themselves of poor quality. Interestingly, recent work has found that adaptation performance is not purely determined by data volume \cite{xiao2026continual,xiao2026dama}, suggesting that data quality may play an equally, if not more, consequential role. While model adaptation is fundamentally bounded by the quality of the data on which it relies, it is therefore essential to systematically investigate data quality as a factor shaping adaptation performance, and to leverage data-centric insights to better understand and improve ASR for Pacific Indigenous languages \cite{gao2023human}.

We formalize data quality in Pacific Indigenous ASR as multi-dimensional, comprising i) \textit{acoustic quality}, which reflects the intelligibility and signal fidelity of the recording under variable recording conditions, ii) \textit{transcript stability}, which captures visible transcription irregularities in the written target, and iii) \textit{audio--text alignment quality}, which reflects whether the recording and transcript correspond at the utterance and word levels. Prior work has studied acoustic quality via reference-less speech intelligibility estimation \cite{kumar2023squim,taal2010stoi}, transcript quality in field-linguistic corpora \cite{leferrand2025field}, and audio--text alignment using multilingual resources that make broad forced-alignment diagnostics feasible \cite{pratap2023mms}. However, these signals are usually considered separately, reflecting only partial data quality. Moreover, for Pacific Indigenous ASR adaptation, identifying the relevant data-quality dimension is important for three reasons. First, from a data-centric perspective, it helps identify potential directions for data collection and guide the construction of a more suitable dataset under the same resource constraint. Second, it helps examine how data quality affects adaptation performance and informs the design of effective PEFT methods that utilize all valuable data. Third, for training methods such as sample reweighting, data quality can serve as a core metric for reweighting training examples and improving adaptation performance.
Curriculum learning (CL) serves as a highly effective paradigm for domain adaptation, particularly when the training corpus exhibits high data quality heterogeneity~\cite{bengio2009curriculum,braun2017accan}.\td{ref} \ys{added.} 
By progressively exposing the model to samples of increasing complexity, CL smooths the optimization landscape and prevents premature convergence to poor local minima caused by noisy or out-of-domain utterances~\cite{bengio2009curriculum}. 
\textit{Rather than scheduling an easy-to-hard ordering, we adopt the reweighting instantiation of this idea:} we translate estimated per-sample difficulty into fixed loss weights computed before training~\cite{kuznetsova2022curriculum,karakasidis2022criteria}, keeping all utterances but scaling their contributions, which provides a natural basis for incorporating data-quality evidence into ASR adaptation.
For Pacific Indigenous languages, model-side difficulty estimation alone is insufficient because high loss can be ambiguous. Degraded acoustic quality and misaligned transcriptions may both lead to high loss, but this often reflects noisy supervision, where the data may be harmful rather than representing learnable training difficulty. Therefore, combining data-side reliability with model-side difficulty estimates could provide a clearer and more effective basis for quality-aware reweighting in Pacific Indigenous ASR.

In this work, we propose QuaSR (Quality-aware sample reweighting), a multi-dimensional data quality scoring framework for Pacific Indigenous ASR. It further leverages the quality scores with model-based difficulty estimates for improved ASR adaptation performance. Specifically, QuaSR quantifies data quality across three complementary dimensions and merges them into a single composite quality score. We further calibrate the model-based difficulty score with this composite score, mitigating cases where model-based scores alone conflate noisy supervision with genuinely informative difficulty. The score therefore serves a dual role: it provides a data-quality diagnostic for analyzing adaptation behavior beyond data volume, and it provides a calibration signal for sample weighting.

QuaSR is validated across four Pacific Indigenous languages: Bislama, Nafsan, Lelepa, and Nguna, and the proposed quality scores successfully interpret adaptation performance, demonstrating their utility as a practical metric for data quality that can inform future data collection efforts.
Furthermore, applying QuaSR yields the largest gains on Lelepa and Nguna, reducing WER by up to 4.01 absolute points and CER by up to 3.76 absolute points.
In-depth analysis additionally reveals that the proposed scores are indicative of sample quality in the latent space.
The contributions of this work are summarised as follows:

\begin{itemize}
    \item This is the first study to systematically analyze data quality for Pacific Indigenous languages in the context of ASR adaptation.
    \item QuaSR provides the first integration of data-centric learning principles into a sample reweighting scheme for ASR adaptation of Pacific Indigenous languages.
    \item Experimental results across four languages demonstrate the effectiveness of QuaSR in the adaptation process, providing a generalized scoring framework applicable to other low-resource languages.
\end{itemize}
\section{Related Work}
\subsection{ASR Adaptation for Pacific Indigenous Languages}

Recent work has begun adapting large pretrained speech models to Pacific Indigenous languages, demonstrating that fine-tuning is feasible even under severely limited data conditions \cite{xiao2026continual,xiao2026dama}. However, recognition performance varies considerably across languages even under comparable model configurations, and the sources of this variability remain poorly understood \cite{ambikairajah2025similarities,dang2025dharawal,xiao2026continual,xiao2026dama}. Existing studies primarily attribute performance differences to data volume and language distance, yet neither factor fully explains the observed gaps. A critical missing piece is data-quality evidence: understanding how acoustic conditions, transcription reliability, and audio–text alignment jointly shape adaptation outcomes, and how such evidence should inform the training process.

\subsection{Sample Difficulty and Training Dynamics}
The notion of sample difficulty has been widely used in quality-aware training like curriculum learning, which proposes ordering training examples from easier to harder, with difficulty approximated in ASR through proxies such as utterance duration, training loss, and model uncertainty \cite{bengio2009curriculum,kuznetsova2022curriculum,karakasidis2022criteria}. Training-dynamics approaches, such as dataset cartography \cite{swayamdipta2020cartography}, extend this by characterizing examples through model behavior across training epochs rather than static metadata alone, revealing whether examples are consistently easy, hard, or unstable for the model. However, for noisy speech corpora such as those found in Pacific Indigenous language collections, high training loss is inherently ambiguous: it may reflect genuinely learnable linguistic difficulty or, equally, unreliable supervision arising from poor acoustics, inconsistent transcription, or misaligned audio–text pairs. This ambiguity motivates sample-weighting criteria that explicitly distinguish model-side learnability from data-side supervision reliability — precisely the gap that QuaSR addresses.

\subsection{Speech Corpus Quality}

Speech corpus quality is multi-dimensional, but Pacific Indigenous ASR makes this issue especially acute. First, Pacific languages (e.g., diverse Austronesian and Papuan families~\cite{lynch2002oceanic,palmer2017newguinea}) are not merely low-resource; they are distributionally highly distant from the Indo-European languages dominating modern speech foundation models~\cite{babu2022xlsr,storey2024languagebias}\td{ref}\ys{added.}. 
Second, Pacific corpora are collected from multiple legacy field linguistics and ethnographic archives.  The audio suffers from unstructured environmental degradation and
conversational overlap~\cite{jones2024cakavian}, while the transcriptions exhibit orthographic fluidity \cite{leferrand2025field}.

Commonly used 
acoustic intelligibility estimators such as SQUIM and STOI provide a practical means of approximating acoustic quality \cite{kumar2023squim,taal2010stoi}. 
\ysrev{Transcript quality is commonly evaluated through transcription consistency or reference-based error measures when repeated annotations or reference transcripts are available~\cite{glenn2010transcription,gao2023human}.} \td{XX are generally used for transcription quality [ref]. } 
Audio--text alignment constitutes a third dimension, which measures 
whether a transcript and its corresponding recording are mismatched 
or poorly segmented at the utterance level, and is generally 
estimated using forced-alignment tools that compute frame-level 
correspondence between the acoustic signal and its transcript 
\cite{pratap2023mms}. Multilingual resources such as MMS make such 
forced-alignment diagnostics feasible across many low-resource 
languages \cite{pratap2023mms}. Whilst each of these dimensions has been studied independently and 
provides useful evidence in its own right, they are rarely 
considered jointly, leaving a gap in our understanding of how their 
combination shapes data reliability. More importantly, how these 
dimensions can be collectively applied to estimate data quality for 
Pacific Indigenous languages, and subsequently leveraged to improve 
ASR adaptation, remains an open question. 
These three dimensions together motivate a more comprehensive and 
principled picture of data quality for low-resource ASR adaptation.

\begin{table}[t!]
\centering
\caption{Data statistics across different languages.}
\vspace{-1mm}
\label{tab:duration}
\small
\resizebox{0.6\columnwidth}{!}{%
\begin{tabular}{@{\extracolsep{\fill}}lccr@{}}
\toprule
Language & Samples & Duration (H) & Avg. (S) \\
\midrule
Bislama & 11,507 & 13.36 & 4.18 \\
Nafsan & 9,226 & 14.71 & 5.74 \\
Lelepa & 3,022 & 3.55 & 4.23 \\
Nguna & 1,115 & 1.02 & 3.29 \\
\midrule
Total & 23,870 & 32.64 & 4.92 \\
\bottomrule
\end{tabular}}
\vspace{-6mm}
\end{table}

\section{Pacific Indigenous Speech Corpus}


We use four labeled corpora derived from records curated by the Pacific and Regional Archive for Digital Sources in Endangered Cultures (PARADISEC): Bislama \cite{thieberger2023bislama}, Nafsan \cite{paradisec1995south_efate}, Lelepa \cite{lacrampe2017lelepa}, and Nguna \cite{facey1988nguna}. 
These datasets are annotated by community experts, providing valuable resources with transcribed data for ASR.
\td{These datasets are annotated by XX, providing valuable resources with transcribed data for ASR.}
\ys{@Austin, sorry, I don't know these datasets are annotated by whom}
Bislama is an English-lexified creole and one of Vanuatu's official languages; Nafsan, Lelepa, and Nguna are Oceanic languages spoken in different regions of Vanuatu.
The corpora differ in duration, elicitation style, recording condition, and transcript consistency, making them a useful testbed for data-quality-aware adaptation. 
Table~\ref{tab:duration} summarizes the corpus size and duration used in this study.
\ysrev{The corpora span 1.02--14.71 hours and 1,115--11,507 utterances. Nafsan has the longest total duration, with 14.71 hours of speech, while Bislama contains the largest number of utterances, with 11,507 samples. Nguna is the smallest corpus, containing 1.02 hours of speech and 1,115 samples. Average utterance duration is comparable across the four languages, ranging from 3.29 to 5.74 seconds, which indicates that most samples are short utterances suitable for ASR training.}
\td{Which one occupies the most data, and which one was collected using only 1 hour of data. The average duration is comparable across four languages, consisting of a few words or a single utterance suitable for ASR.}

%


\section{QuaSR: Quality-Aware Sample Reweighting}
\label{sec:method}

QuaSR defines sample utility as the product of data-side reliability and model-side learnability. 
The data-side reliability term combines three corpus-quality signals: acoustic quality ($Q$), transcript stability ($T$), and audio--text alignment reliability ($A$). 
\ysrev{The model-side learnability term is estimated from sample-level loss values recorded during an unweighted baseline training run.} \td{i.e., loss values?}
QuaSR multiplies these two terms so that a sample receives high training influence only when it is both reliable in data quality and learnable for the current model. The resulting sample utility is then 
incorporated in a soft loss weight for 
LoRA adaptation.


\subsection{Data Reliability}

Data-side reliability estimates whether an utterance is a high-quality sample suitable for reliable supervision by capturing three complementary dimensions. Acoustic quality ($Q$), estimated using the SQUIM estimate of short-time objective intelligibility (STOI), measures whether the speech signal is sufficiently intelligible for ASR training \cite{kumar2023squim,taal2010stoi}.
Transcript stability ($T$) measures visible transcription irregularities in the transcript, including markers of incomplete transcription, digit characters, and isolated non-lexical symbols.
Audio--text alignment reliability ($A$), estimated from MMS forced-alignment diagnostics, combines word-level alignment confidence with utterance-level CTC transcript agreement to measure whether the transcript is well matched to the recording \cite{pratap2023mms}.


\ysrev{Acoustic quality is measured using the reference-less SQUIM estimate of STOI~\cite{kumar2023squim,taal2010stoi}. For each utterance, the pretrained SQUIM model takes the input waveform and predicts an STOI score without requiring a clean reference signal. We use this predicted STOI score as the raw acoustic-quality score $\tilde{Q}_i$.}
Low values indicate speech that is likely difficult to perceive because of noise, channel mismatch, or other acoustic degradation, even when the transcript itself is valid.
This component therefore estimates the acoustic reliability of the recording as a supervisory signal.\td{what is the Qi and $\pho$? do we need both? here we mainly want to show how to calculate the squim. that's all.}
\td{do you have an equation for si, here it is just a symbol and does not show any useful information.}
\ys{removed the uninformative equation because si has no exact equation.}


Transcript stability penalizes visible transcription irregularities in the transcript without imposing a language-specific dictionary.
\vspace{-3mm}
\begin{equation}
\tilde{T}_i=1-\min\left(1,\frac{d_i+e_i+u_i}{n_i}\right), 
\end{equation}
where $d_i$, $e_i$, and $u_i$ count digit characters, markers of incomplete transcription, and isolated non-lexical symbols, respectively, and $n_i$ is the number of non-space characters.
A lower $\tilde{T}_i$ therefore indicates that these counted irregularities occupy a larger fraction of the transcript.
Unicode letters and combining marks are not penalized, so the score does not treat Indigenous orthography as noise.


\ysrev{Audio--text alignment reliability is computed from MMS forced-alignment diagnostics~\cite{pratap2023mms}. For utterance $i$, MMS aligns the input waveform with the normalized transcript. Let $J_i$ be the number of transcript words. For word $j$, let $\mathcal{S}_{ij}$ denote its aligned token spans, $c_{ijk}$ the MMS confidence score of token span $k$, and $\delta_{ij}$ the aligned word duration. We first compute the word-level confidence as}
\begin{equation}
q_{ij}=\frac{1}{|\mathcal{S}_{ij}|}\sum_{k=1}^{|\mathcal{S}_{ij}|} c_{ijk},
\quad
m_i^{\mathrm{word}}=
\frac{\sum_{j=1}^{J_i}\delta_{ij}q_{ij}}
{\sum_{j=1}^{J_i}\delta_{ij}}.
\end{equation}
\ysrev{Let $\hat{y}_i$ denote the text obtained by greedy CTC decoding from the MMS emissions, and let $\bar{y}_i$ denote the normalized reference transcript. We compare these two texts using character error rate (CER):}
\begin{equation}
m_i^{\mathrm{ctc}}=\max\left(0,1-\operatorname{CER}(\bar{y}_i,\hat{y}_i)\right).
\end{equation}
\ysrev{The raw audio--text alignment reliability score is then}
\begin{equation}
\tilde{A}_i=\frac{1}{2}m_i^{\mathrm{word}}+\frac{1}{2}m_i^{\mathrm{ctc}}.
\end{equation}
\td{again here the m is just a symbol and not clear how it is calculated from the speech and transcrition. you need to provide the real equation to claucate m. Also what is pho here?}
\ys{pho is moved to later}

Together, these diagnostics assign each training utterance an audio--text alignment reliability score: low values indicate that the transcript may be poorly matched to the recording, either because word-level alignments have low confidence or because the utterance-level transcript agreement is low.


We rank-normalize each score for each language using $\rho_l(\cdot)$ before combining components.\td{until here you mention about normalisation, so before this, no need to mention normalization yet, just discuss how it is calculated. }
\begin{equation}
Q_i=\rho_l(\tilde{Q}_i), \quad
T_i=\rho_l(\tilde{T}_i), \quad
A_i=\frac{1}{2}\rho_l(m_i^{\mathrm{word}})+\frac{1}{2}\rho_l(m_i^{\mathrm{ctc}}).
\end{equation}
This step converts each diagnostic into a relative component score: within a given language, utterances with better raw diagnostic values receive higher normalized component scores.
We use ranks because the raw diagnostics have different numerical scales and cannot be combined directly; for example, SQUIM/STOI scores, transcript-irregularity counts, and MMS alignment confidences are not comparable as raw quantities.


We evaluate four fixed reliability variants that combine the normalized components:
\begin{align}
R_i^{AT}  &= (A_i+T_i)/2,       &
R_i^{AQ}  &= (A_i+Q_i)/2,       \nonumber\\
R_i^{TQ}  &= (T_i+Q_i)/2,       &
R_i^{ATQ} &= (A_i+T_i+Q_i)/3.
\end{align}
The equal weights make each variant a controlled diagnostic combination: included components contribute symmetrically, and variants differ only in which reliability signals they contain.
This avoids tuning component weights on very small development sets.
The variants test which type of data-side reliability evidence is most useful for each language: transcript--alignment ($AT$), acoustic--alignment ($AQ$), transcript--acoustic ($TQ$), or all three signals ($ATQ$).

%
%



\subsection{Model-Side Learnability}
\ysrev{We estimate model-side learnability using sample-level losses recorded during an unweighted baseline training run.
The intuition is that samples consistently assigned low loss are easier for the model to learn, whereas samples whose losses remain high or fluctuate substantially are more difficult to learn.
Specifically, we use the teacher-forced decoder loss as our measure~\cite{swayamdipta2020cartography}.
Teacher forcing is a standard training technique in which the model is given the correct transcript one word at a time and asked to predict the next token.
This allows us to measure how confidently the model predicts each word in the transcript, given the audio and all preceding correct tokens.
Averaging over target tokens prevents longer utterances from receiving larger losses simply because they contain more tokens.}

\td{we estimate model-side learnability using the sample-level losses recorded during an unweighted baseline training run. The key intuition is that samples that are consistently assigned low loss are easier for the model to learn, whereas samples whose losses remain high or fluctuate substantially are more difficult to learn. Specifically, we use the teacher-forced decoder loss as our measure~\cite{swayamdipta2020cartography}. 
Teacher forcing is a standard training technique in which 
the model is given the correct transcript one word at a time and 
asked to predict the next word. This allows us to measure how 
confidently the model predicts each word in the transcript, given 
the audio and all preceding correct words. The loss is averaged 
over all transcript tokens so that longer utterances are not 
penalized simply for being longer.}

\ysrev{Formally, given an utterance--transcript pair $(x_i, y_i)$, we define $\ell_{i,e}$ as the token-averaged teacher-forced decoder loss for sample $i$ at epoch $e$:}

\td{Formally, given an utterance--transcript pair $(x_i, y_i)$, we 
define $\ell_{i,e}$ as the token-averaged teacher-forced decoder 
loss for sample $i$ at epoch $e$:}
\begin{equation}
\ell_{i,e}
=
-\frac{1}{|M_i|}
\sum_{t \in M_i}
\log p_{\theta_{i,e}}(y_{i,t} \mid y_{i,<t}, x_i),
\end{equation}
where $M_i$ is the set of target token positions, 
$p_{\theta_{i,e}}(y_{i,t} \mid y_{i,<t}, x_i)$ is the 
probability the model assigns to the correct token $y_{i,t}$ 
given the audio $x_i$ and all preceding correct tokens 
$y_{i,<t}$, and $\theta_{i,e}$ denotes the model parameters 
at epoch $e$ when sample $i$ is observed.


\ysrev{We convert the per-epoch losses into a bounded learnability score by averaging the exponentiated negative loss across training epochs:}

\td{We convert the per-epoch losses into a bounded learnability score by averaging the exponentiated negative loss across training epochs:}
\vspace{-3mm}
\begin{equation}
C_i=\frac{1}{E}\sum_{e=1}^{E}\exp(-\ell_{i,e}).
\end{equation}

\ysrev{We empirically choose $E=16$ epochs in all experiments. A higher $C_i$ indicates that the baseline model consistently assigns low loss to sample $i$ throughout training, suggesting that it is more readily learned. Conversely, persistently high loss corresponds to lower learnability for the current adaptation model, rather than implying that the utterance is unusable or incorrectly transcribed.}

\td{We empirically choose $E=16$ epochs in all experiments. A higher $C_i$ indicates that the baseline model consistently assigns low loss to sample $i$ throughout training, suggesting that it is more readily learned. Conversely, persistently high loss corresponds to lower learnability for the current adaptation model, rather than implying that the utterance is unusable or incorrectly transcribed.}

\ysrev{Since sample weighting and weight normalisation are performed independently for each language, $C_i$ serves as a relative learnability measure within a language and is not intended for comparisons across languages. For each weighted adaptation run, the $C_i$ values are computed once from the corresponding unweighted baseline using the same random seed and remain fixed throughout training.}

\td{Since sample weighting and weight normalisation are performed independently for each language, $C_i$ serves as a relative learnability measure within a language and is not intended for comparisons across languages. For each weighted adaptation run, the $C_i$ values are computed once from the corresponding unweighted baseline using the same random seed and remain fixed throughout training.}


\subsection{Proposed QuaSR}

\ysrev{The proposed QuaSR combines data-side reliability and model-side learnability to estimate the overall utility of each training sample. Samples that are both reliable as supervision and readily learnable by the adaptation model are assigned larger weights, while samples that are unreliable, difficult to learn, or both receive smaller weights.}

\td{The proposed QuaSR combines data-side reliability and model-side learnability to estimate the overall utility of each training sample. Samples that are both reliable as supervision and readily learnable by the adaptation model are assigned larger weights, while samples that are unreliable, difficult to learn, or both receive smaller weights.}

\ysrev{Let $\mathcal{I}_l$ denote the set of training utterances for language $l$, where $N_l=|\mathcal{I}_l|$. For each reliability variant $v$, the sample utility score $U_i^{(v)}$ is defined as:}
\td{Let $\mathcal{I}_l$ denote the set of training utterances for language $l$, where $N_l=|\mathcal{I}_l|$. For each reliability variant $v$, the sample utility score $U_i^{(v)}$ is defined as: }
\begin{equation}
U_i^{(v)} = R_i^{(v)}C_i,
\end{equation}
where $R_i^{(v)}$ is the data-side reliability score and $C_i$ is the model-side learnability score. The multiplicative formulation acts as a conservative gating mechanism, assigning a high weight only when a sample is both reliable and learnable. 

\ysrev{The utility scores are then mean-normalized within each language to obtain the training weights:}

\td{The utility scores are then mean-normalized within each language to obtain the training weights: }
\begin{equation}
w_i^{(v)}=\frac{U_i^{(v)}}{\frac{1}{N_l}\sum_{j\in \mathcal{I}_l}U_j^{(v)}}.
\end{equation}
Mean normalization ensures that the average sample weight within each language is one, preserving the overall scale of the training objective and enabling fair comparisons across different weighting strategies.


\ysrev{During adaptation, the normalized weight $w_i^{(v)}$ is applied uniformly to all target-token losses for sample $i$. Given a minibatch $\mathcal{B}$, the weighted training objective is:}

\td{During adaptation, the normalized weight $w_i^{(v)}$ is applied uniformly to all target-token losses for sample $i$. Given a minibatch $\mathcal{B}$, the weighted training objective is:}
\vspace{-3mm}
\begin{equation}
\mathcal{L}_{\mathcal{B}}^{(v)}(\theta)
=
\frac{
\sum_{i\in\mathcal{B}} w_i^{(v)}
\sum_{t\in M_i}
-\log p_{\theta}(y_{i,t}\mid y_{i,<t},x_i)
}{
\sum_{i\in\mathcal{B}} w_i^{(v)} |M_i|
},
\end{equation}
where $M_i$ denotes the target-token positions included in the teacher-forced loss.
The weighting is soft in the sense that all samples are retained in the training set, while lower-utility samples receive smaller loss weights. 

\section{Experiments Setup}
\label{sec:experiments}


To make the reliability variants comparable, we keep the adaptation configuration fixed across conditions.
For each language, we use a fixed utterance-level 80/10/10 train/development/test split. 
All runs fine-tune Whisper-small with LoRA adapters on both encoder and decoder modules.
Encoder adapters target the self-attention $q$, $k$, $v$, and output projection modules; decoder adapters target the same attention projections and the feed-forward $fc_1$ and $fc_2$ modules. 
Training uses LoRA rank 64, alpha 64, dropout 0, no bias adaptation, AdamW, peak learning rate $10^{-4}$, train and validation batch size 16, eight workers, random sorting, 16 epochs, and cosine decay. 
Warm-up uses 500, 370, 120, and 45 steps for Bislama, Nafsan, Lelepa, and Nguna, respectively. 

For each language, three unweighted seeds are used and the mean and standard deviations are reported. The performance is reported using WER (word error rate) and CER (character error rate). 


\section{Results and Analysis}
\label{sec:results-analysis}


\subsection{Baseline Adaptation Performance}

\begin{table}[t]
\centering
\caption{Whisper-small LoRA baseline performance.}
\vspace{-2mm}
\label{tab:baseline}
\small
\begin{tabular*}{0.8\columnwidth}{@{\extracolsep{\fill}}lcr@{}}
\toprule
Language & WER & CER \\
\midrule
Bislama & 21.73 $\pm$ 0.66 & 10.39 $\pm$ 0.39 \\
Nafsan & 50.00 $\pm$ 0.44 & 19.32 $\pm$ 0.63 \\
Lelepa & 69.72 $\pm$ 0.55 & 27.75 $\pm$ 0.72 \\
Nguna & 32.26 $\pm$ 2.07 & 11.43 $\pm$ 0.65 \\
\bottomrule
\end{tabular*}
\vspace{-5mm}
\end{table}

Table \ref{tab:baseline} shows the standard LoRA baseline adaptation performance, aiming to examine the reliability of the proposed data quality scores in explaining standard LoRA performance. Evidently, the standard LoRA performance is not ordered by corpus volume alone, referring to Tables~\ref{tab:duration}. 
Nafsan has the largest training duration but substantially higher error than Bislama, while Nguna has the smallest training set but outperforms Lelepa under the same LoRA configuration.

\begin{figure*}[t]
\centering
\includegraphics[width=0.8\textwidth]{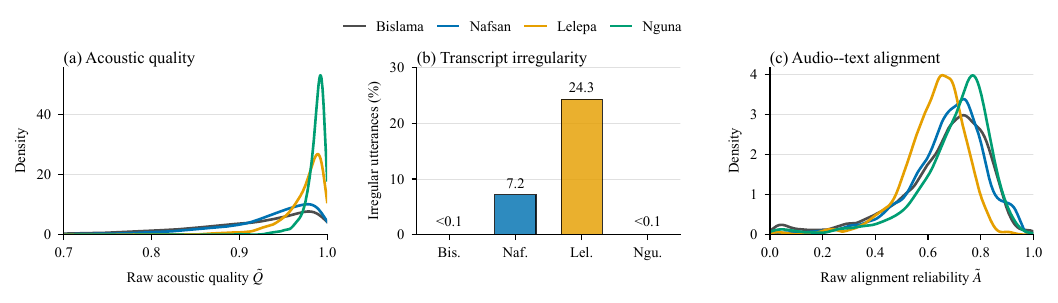}
\vspace{-5mm}
\caption{The data quality comparison across four languages: (a) acoustic quality $\tilde{Q}$, (b) transcript quality measured as the irregularity rate, and (c) audio--text alignment reliability $\tilde{A}$.}
\vspace{-5mm}
\label{fig:raw-quality-diagnostics}
\end{figure*}

We additionally plot the proposed score distributions for all four languages in Figure~\ref{fig:raw-quality-diagnostics}.
Figure~\ref{fig:raw-quality-diagnostics}(a) shows that Bislama and Nafsan include more lower-scoring acoustic samples, while Lelepa and Nguna are concentrated near the high-quality end. 
Figure~\ref{fig:raw-quality-diagnostics}(b) shows that transcript irregularities are rare in Bislama and Nguna, occur in a small subset of Nafsan utterances, and are most frequent in Lelepa. Therefore, Lelepa has the worst transcription quality, which may explain its poor adaptation performance. 
Figure~\ref{fig:raw-quality-diagnostics}(c) indicates that Lelepa has a larger share of lower raw audio--text alignment reliability scores, further validating its worst performance. Nguna has higher alignment reliability. Its high acoustic scores and overall quality may help explain its better adaptation performance, even with the minimum amount of data. These results suggest that data volume is not the key factor in adaptation performance; rather, data quality plays a more important role. The proposed scores are also reliable indicators for evaluating data quality.

\subsection{Quality-Aware Reweighting Results}

\begin{table}[t]
\centering
\caption{Performance comparison of different combinations of data reliability scores. 
Deltas are absolute WER and CER compared to baseline. } 
\vspace{-2mm}
\label{tab:mean-results}
\resizebox{0.9\columnwidth}{!}{%
\begin{tabular}{llrrrr}
\toprule
Language & Variant & WER & $\Delta$WER & CER & $\Delta$CER \\
\midrule
Bislama & Baseline & 21.73 $\pm$ 0.66 & -- & 10.39 $\pm$ 0.39 & -- \\
        & AT  & 21.63 $\pm$ 0.36 & -0.10 & 10.26 $\pm$ 0.20 & -0.14 \\
        & AQ  & 21.87 $\pm$ 0.29 & +0.14 & 10.27 $\pm$ 0.19 & -0.12 \\
        & TQ  & \gain{21.37 $\pm$ 0.36} & \gain{-0.36} & \gain{10.01 $\pm$ 0.11} & \gain{-0.39} \\
        & ATQ & 21.69 $\pm$ 0.44 & -0.04 & 10.32 $\pm$ 0.43 & -0.07 \\
\midrule
Nafsan & Baseline & 50.00 $\pm$ 0.44 & -- & 19.32 $\pm$ 0.63 & -- \\
       & AT  & 49.78 $\pm$ 0.71 & -0.23 & 19.02 $\pm$ 0.26 & -0.30 \\
       & AQ  & 50.37 $\pm$ 1.74 & +0.37 & 19.87 $\pm$ 1.05 & +0.55 \\
       & TQ  & \gain{49.55 $\pm$ 1.08} & \gain{-0.45} & \gain{18.95 $\pm$ 0.54} & \gain{-0.37} \\
       & ATQ & 50.16 $\pm$ 0.91 & +0.16 & 19.38 $\pm$ 0.63 & +0.06 \\
\midrule
Lelepa & Baseline & 69.72 $\pm$ 0.55 & -- & 27.75 $\pm$ 0.72 & -- \\
       & AT  & \gain{66.17 $\pm$ 1.52} & \gain{-3.55} & \gain{23.99 $\pm$ 1.19} & \gain{-3.76} \\
       & AQ  & 67.80 $\pm$ 1.77 & -1.92 & 25.56 $\pm$ 1.23 & -2.19 \\
       & TQ  & 67.43 $\pm$ 1.55 & -2.29 & 25.25 $\pm$ 1.12 & -2.50 \\
       & ATQ & 68.07 $\pm$ 0.99 & -1.65 & 25.86 $\pm$ 0.83 & -1.89 \\
\midrule
Nguna & Baseline & 32.26 $\pm$ 2.07 & -- & 11.43 $\pm$ 0.65 & -- \\
      & AT  & 31.28 $\pm$ 0.88 & -0.99 & 11.02 $\pm$ 0.38 & -0.41 \\
      & AQ  & \gain{28.25 $\pm$ 1.49} & \gain{-4.01} & \gain{9.02 $\pm$ 0.41} & \gain{-2.41} \\
      & TQ  & 32.09 $\pm$ 1.55 & -0.18 & 11.78 $\pm$ 0.87 & +0.35 \\
      & ATQ & 31.51 $\pm$ 0.62 & -0.75 & 11.07 $\pm$ 0.30 & -0.36 \\
\bottomrule
\end{tabular}%
}
\vspace{-6mm}
\end{table}


Table~\ref{tab:mean-results} reports the performance using QuaSR 
for the four languages.
QuaSR shows the best-performing reliability variant reduces both WER and CER relative to the standard LoRA baseline, with the largest gains observed for Lelepa and Nguna.
Across languages, the strongest weighted variants reduce WER by up to 12.4\% and CER by up to 21.1\% relatively. 
This suggests that QuaSR can generalize across different languages. 


The comparison of different combinations of data reliability scores shows that the best-performing reliability score varies across languages, and no single variant consistently dominates across all languages. For Lelepa, transcript and alignment reliability ($AT$) gives the largest improvement, reducing WER by 3.55 and CER by 3.76, corresponding to relative reductions of 5.1\% and 13.5\%, respectively.
This pattern is consistent with data quality scores: Lelepa shows the highest transcript-irregularity rate in Figure~\ref{fig:raw-quality-diagnostics}(b) and weaker audio--text alignment reliability in Figure~\ref{fig:raw-quality-diagnostics}(c), while its acoustic-quality distribution is relatively strong in Figure~\ref{fig:raw-quality-diagnostics}(a). 


The smaller gain from $ATQ$ suggests that adding acoustic quality is less useful for Lelepa than emphasizing transcript stability and audio-text alignment. For Nguna, the combination of acoustic and alignment reliability ($AQ$) shows the largest improvement, reducing WER by 4.01 and CER by 2.41, corresponding to relative reductions of 12.4\% and 21.1\%, respectively. 
\ysrev{This is consistent with Figure~\ref{fig:raw-quality-diagnostics}(a)--(c): Nguna has almost no transcript irregularities, making transcript stability less informative, while acoustic quality and audio--text alignment provide more useful weighting signals.} \td{since you change the figure layout, make sure the number you refer to the figure are now updated accordingly. }


Bislama and Nafsan show smaller but still positive improvements. For Bislama, $TQ$ reduces WER by 0.36 and CER by 0.39, corresponding to relative reductions of 1.7\% and 3.8\%, respectively.
For Bislama, the standard LoRA baseline is already relatively strong and the data quality scores indicate comparatively high data quality, so QuaSR yields a smaller but still consistent improvement. 


\subsection{Latent Space Visualization}

\begin{figure}[t]
\centering
\vspace{-6mm}
\includegraphics[width=\columnwidth]{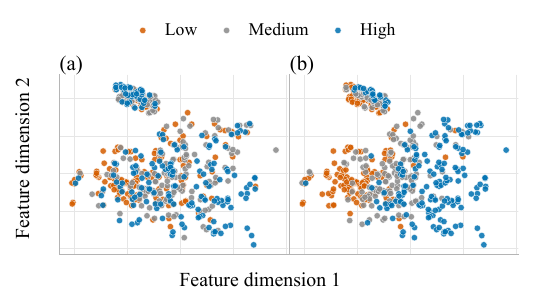}
\vspace{-6mm}
\caption{Comparison of t-SNE feature representations for Lelepa using (a) model-side learnability ($C_i$) only and (b) the proposed utility-based scoring \(U_i^{(AT)} = R_i^{AT} C_i\).}
\vspace{-2mm}
\label{fig:lelepa-tsne}
\end{figure}

\ysrev{To further validate the proposed quality score, which incorporates both data reliability and model learnability, we categorize samples in the latent space into low, medium, and high difficulty buckets using two criteria: model learnability alone ($C_i$) and the proposed utility score $U_i^{(AT)} = R_i^{AT} C_i$. As shown in Figure~\ref{fig:lelepa-tsne} for Lelepa, when the best-performing transcript--alignment reliability score is incorporated into the proposed score, approximately 41--51\% of samples shift into a different difficulty group in Figure~\ref{fig:lelepa-tsne}(b), compared to using model learnability alone in Figure~\ref{fig:lelepa-tsne}(a). This confirms that transcript--alignment reliability carries independent and meaningful information about sample utility: a sample that appears difficult to learn from may in fact be unreliably transcribed rather than genuinely hard, and vice versa. This suggests that the proposed data reliability scores can effectively capture true data quality, thereby enhancing the representation in the latent space and, in turn, improving model learning. This is consistently observed for all other three languages as well.}

\td{can we add one last sentence here: This is consistently observed for all other three languages as well. } \ys{yes, it is true} 



\subsection{Comparison to Hard Filtering}

\begin{table}[t]
\centering

\caption{Comparison of QuaSR to hard filtering for Lelepa and Nguna.}
\label{tab:hard-filter}
\small
\resizebox{0.8\columnwidth}{!}{%
\begin{tabular}{@{\extracolsep{\fill}}llrr@{}}
\toprule
Language & Configuration & WER & CER \\
\midrule
Lelepa & Baseline & 69.72 $\pm$ 0.55 & 27.75 $\pm$ 0.72 \\
       & $AT$ soft & \gain{66.17 $\pm$ 1.52} & \gain{23.99 $\pm$ 1.19} \\
       & $AT$ filter 10\% & 70.13 $\pm$ 1.20 & 27.87 $\pm$ 0.62 \\
       & $AT$ filter 20\% & 72.11 $\pm$ 0.84 & 29.22 $\pm$ 0.78 \\
\midrule
Nguna  & Baseline & 32.26 $\pm$ 2.07 & 11.43 $\pm$ 0.65 \\
       & $AQ$ soft & \gain{28.25 $\pm$ 1.49} & \gain{9.02 $\pm$ 0.41} \\
       & $AQ$ filter 10\% & 32.58 $\pm$ 0.83 & 11.73 $\pm$ 0.16 \\
       & $AQ$ filter 20\% & 42.10 $\pm$ 7.28 & 17.58 $\pm$ 5.03 \\
\bottomrule
\end{tabular}}
\vspace{-5mm}
\end{table}

Table~\ref{tab:hard-filter} further compares QuaSR with existing hard filtering methods, which removes the most difficult samples based on the proposed difficulty criterion instead of soft reweighting (as in QuaSR).  
We focus the hard-filtering comparison on the two smallest corpora, Lelepa (3.55 h) and Nguna (1.02 h). Hard filtering further removes the lowest-ranked 10\% or 20\% of training samples under the corresponding reliability variant, with development and test sets kept unchanged.
In both comparisons for Lelepa and Nguna, QuaSR yields lower mean WER and CER than hard-filtering method. 
The comparison suggests that the lowest-ranked utterances are not uniformly harmful in these small corpora.
QuaSR can reduce their influence while still preserving useful acoustic and linguistic evidence. 
This is particularly important for low-resource ASR, where the available training data are limited.
\td{why don't we have the results for the  other two lanagues? needs a justification. You cannot say that becasue these two are best performed, so we analyse this two. Thne the question will be: does that mean your method works worse than hard filtering for the other two languages? So we need a stronger justificaiton. }
\ys{@Austin, sorry here i have no idea for the stronger justification why we choose the two.}

\section{Conclusion}
We propose QuaSR, a Quality-Aware Sample Reweighting framework for Pacific-language ASR that separates data reliability from model learnability to score sample utility. Combining data quality with model dynamics outperforms learnability-based weighting alone, and the most useful reliability evidence proves language-dependent: transcript–alignment signals help Lelepa, acoustic–alignment signals help Nguna, while Nafsan remains limited by corpus heterogeneity. This argues for language-specific reliability signals over a single universal difficulty measure in quality-aware training for under-served, low-resource languages.

\section{Generative AI Use Disclosure}

Generative AI tools were used only to polish the language of this manuscript, such as grammar and wording.

\bibliographystyle{IEEEtran}
\bibliography{references}

\end{document}